\documentclass[prd,twocolumn,nofootinbib]{revtex4}

\usepackage[utf8]{inputenc} % allow utf-8 input
\usepackage[T1]{fontenc}    % use 8-bit T1 fonts
\usepackage{hyperref}       % hyperlinks
\usepackage{url}            % simple URL typesetting
\usepackage{amsfonts}       % blackboard math symbols
\usepackage{nicefrac}       % compact symbols for 1/2, etc.
\usepackage{microtype}      % microtypography
\usepackage{graphicx}       % graphics
\graphicspath{{media/}}     % organize your images and other figures under media/ folder
\usepackage{amsmath}
\usepackage{feynmp-auto}
\usepackage{amssymb}
\usepackage{ragged2e}
\usepackage{blindtext}
\usepackage{float}
\usepackage{placeins}
\usepackage{float}
\usepackage{xcolor}
\usepackage{slashed}

\begin{document}

\title{Implications of recent $\left(g-2\right)_{\mu}$ measurements for MeV-GeV dark sector searches}

\author{Aleksandr Pustyntsev}
\affiliation{Institut f\"ur Kernphysik and $\text{PRISMA}^+$ Cluster of Excellence, Johannes Gutenberg Universit\"at, D-55099 Mainz, Germany}
\author{Marc Vanderhaeghen}
\affiliation{Institut f\"ur Kernphysik and $\text{PRISMA}^+$ Cluster of Excellence, Johannes Gutenberg Universit\"at, D-55099 Mainz, Germany}

\date{\today}

\begin{abstract}
Recent theoretical and experimental studies of the muon magnetic moment indicate the absence of the previously reported discrepancy, providing a vital opportunity to constrain potential BSM physics. In this work, we explore the MeV-GeV mass range, where existing exclusion limits remain relatively loose. We analyze both scalar and pseudoscalar as well as vector and axial vector mediators. We demonstrate that the new bounds are not only comparable to - but in several cases, significantly more stringent than - the constraints obtained from previous collider experiments, even when near-future projections are considered.
\end{abstract}

\maketitle

\section{Introduction}\label{sec1}

The Muon $g-2$ collaboration at Fermilab has recently released its final measurement of the muon's anomalous magnetic dipole moment $ a_{\mu, \,\text{exp.}} = 116 592 071.5\left(14.5\right) \times 10^{-11}$ \cite{Muong-2:2025xyk}. With the latest Standard Model (SM) theoretical prediction from \cite{Aliberti:2025beg}, given as $a_{\mu, \,\text{th.}} = 116592033\left(62\right) \times 10^{-11}$, the discrepancy between the experimental findings and the refined SM expectation amounts to $38.5\left(63.7\right) \times 10^{-11}$ and now falls within the $1\sigma$ interval.

Due to the enhanced sensitivity of the muon to high-energy physics, this very fact establishes severe bounds for any potential Beyond Standard Model (BSM) scenario. Axions \cite{Peccei:1977hh, Peccei:1977ur,Weinberg:1977ma, Wilczek:1977pj} and axion-like particles \cite{Jaeckel:2010ni,Baker:2013zta,Graham:2015ouw} (or just ALPs), which have strong theoretical motivation both within the SM and as extensions of the SM, serve as viable dark matter candidates. MeV-GeV mass range is of particular interest, as the experimental constraints in that region remain relatively weak \cite{AxionLimits}.

Another compelling direction of BSM studies is the search for a dark sector and the associated mediators, such as dark photons \cite{Fabbrichesi:2020wbt,Caputo:2021eaa}. Kinetic mixing of such particles with SM photons leads to a non-zero coupling to SM fermions and thus yields  corrections to their magnetic moments. In this work we investigate both vector and axial vector possibilities of such mediators in view of the new constraints provided by the $\left(g-2\right)_{\mu}$ measurement. 

We organize this paper as follows. Section \ref{sec2} provides an overview of the ALP formalism, detailing its couplings, potential corrections to $\left(g-2\right)_{\mu}$, and the resulting constraints. We also include a scalar BSM particle in our analysis. In section \ref{sec3}, we extend our work towards dark photon searches for both possible parities. Section \ref{sec4} provides a summary of our work.

\section{ALP contributions to $\left(g-2\right)_{\mu}$}\label{sec2}

The interaction between ALPs and SM fermions respects the shift symmetry, implying that it can only arise through derivative-like couplings. The corresponding lowest-order Lagrangian takes the form

\begin{equation}\label{eq:der}
\mathcal{L}_{aff} = -\frac{g_{aff}}{2m_f} \partial_{\mu} a \, \Bar{f} \gamma^5 \gamma^{\mu} f,
\end{equation}
with $f$ being the fermion field, $m_f$ is its mass and $g_{aff}$ is the dimensionless coupling constant.

\begin{figure}
\centering
\includegraphics{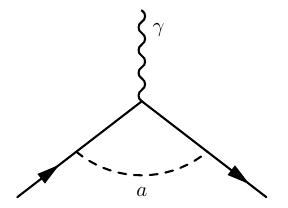} 
\caption{Yukawa-like correction to the anomalous magnetic moment.}
\label{fig:dyuk}
\end{figure}

The Yukawa-like 1-loop contribution to the fermion dipole moment is shown in Fig. \ref{fig:dyuk}. It leads to \cite{Chen:2015vqy}

\begin{equation}\label{eq:yuk}
\Delta a^Y = - \frac{m_f^2 g_{aff}^2}{8\pi^2}  \int_0^1 \frac{\left(1-x\right)^3 }{m_f^2 \left(1-x\right)^2 +m_a^2x}dx.
\end{equation}

\begin{figure}
    \centering
    \includegraphics[width=\linewidth]{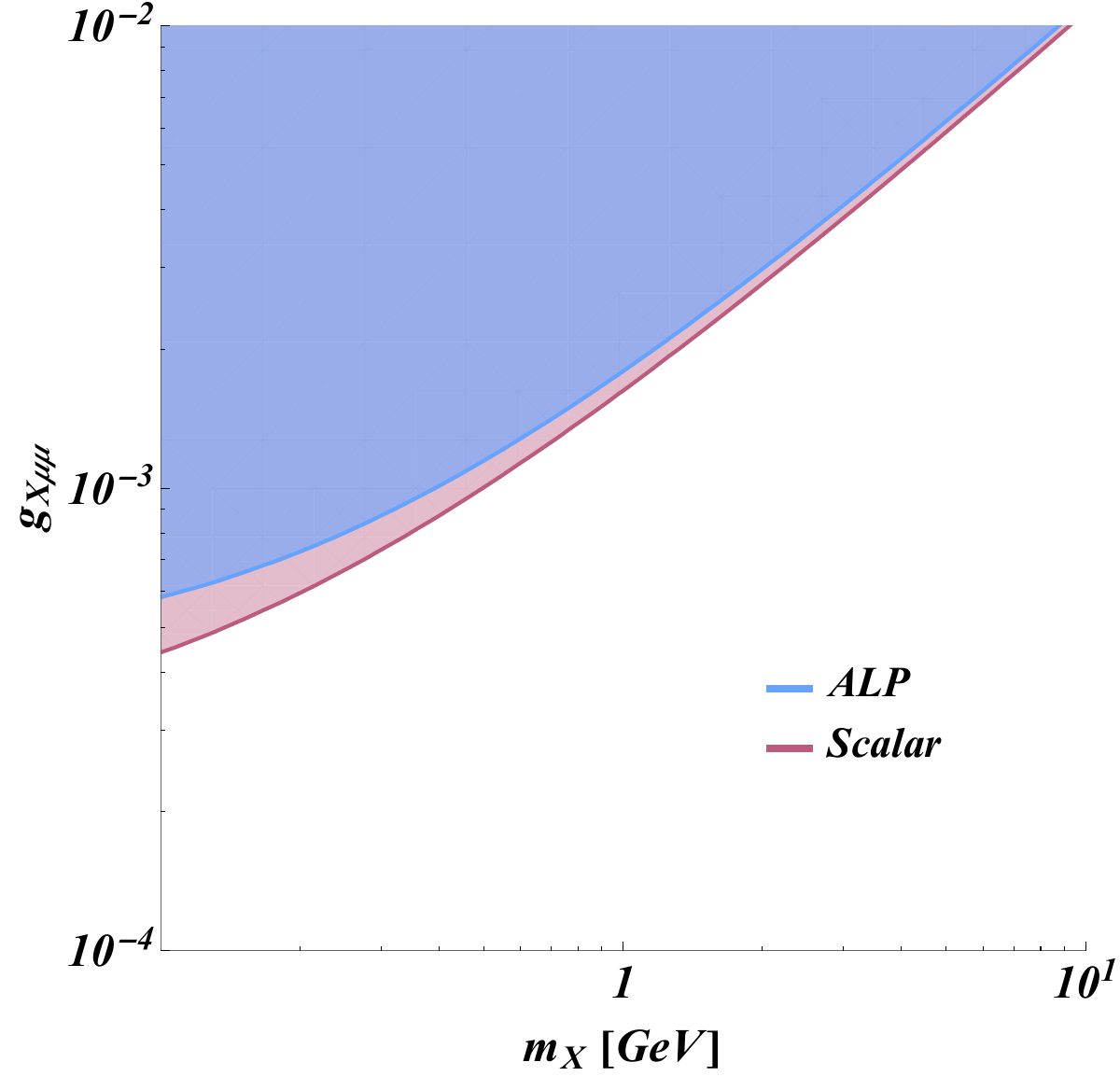}
    \caption{ALP-muon and scalar-muon coupling constraints under the assumption of a single-coupling contribution. $X$ denotes either ALP or scalar. The colored regions are excluded at $2\sigma$ level.}
    \label{fig:LeptonConstr}
\end{figure}

As this expression is negative and the originally reported $\left(g-2\right)_{\mu}$ discrepancy between experiment and theory was positive, this contribution has typically been considered as bearing no significance on its own (i.e., unless there are other contributions that could compensate for it \cite{Marciano:2016yhf}). Given the recent results demonstrate no discrepancy~\cite{Muong-2:2025xyk,Aliberti:2025beg}, we can now use it to constrain the ALP parameter space, as shown in Fig. \ref{fig:LeptonConstr}. 

On another hand, the generic gauge-invariant interaction of ALPs with electroweak sector is given by the Lagrangian \cite{Dolan:2017osp}

\begin{equation}\label{eq:ew}
\mathcal{L}_{aEW} = -\frac{g_{aBB}}{4}aB^{\mu \nu}\tilde{B}_{\mu \nu} -\frac{g_{aWW}}{4}a \textbf{W}^{\mu \nu}\tilde{\textbf{W}}_{\mu \nu},
\end{equation}
where $B^{\mu \nu}$ and $\textbf{W}^{\mu \nu}$ stand for the $U\left(1\right)$ and $SU\left(2\right)$ field tensors, respectively. The dual pseudotensors are $\tilde{B}_{\mu \nu} = \frac{1}{2}\varepsilon_{\mu \nu \lambda \sigma} {B}^{ \lambda \sigma}$ and similarly for $\textbf{W}^{\mu \nu}$. The coupling constants $g_{a BB}$ and $g_{a WW}$ are of $\mbox{GeV}^{-1}$ dimension.

An effective interaction of ALPs and photons is then described by 

\begin{equation}
\mathcal{L}_{a \gamma \gamma} = -\frac{g_{a \gamma \gamma}}{4}aF^{\mu \nu}\tilde{F}_{\mu \nu} - \frac{g_{a \gamma Z}}{2}aF^{\mu \nu}\tilde{Z}_{\mu \nu},
\end{equation}
where the new coupling constants are

\begin{align}
& g_{a\gamma\gamma} = g_{aBB} \cos^2{\theta_w}+g_{aWW} \sin^2{\theta_w}, \\
& g_{a\gamma Z} = \frac{g_{aWW}-g_{aBB}}{2} \sin{\left(2\theta_w\right)},
\end{align}
with $\theta_w$ denoting the weak mixing angle.

\begin{figure}
\centering
\includegraphics{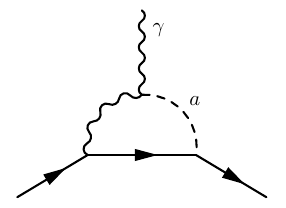} 
\caption{Barr-Zee correction to the anomalous magnetic moment.}
\label{fig:dbrz}
\end{figure}

Experimental searches for flavor-changing processes impose stringent constraints on the coupling $g_{aWW}$~\cite{Izaguirre:2016dfi,BaBar:2021ich}, whereas the limits on $g_{aBB}$ are comparatively more relaxed. Thus, it is reasonable to assume that the ALP coupling to the $\textbf{W}$ boson is at best subdominant, i.e., $g_{aBB} \gg g_{aWW}$. Under this assumption, the two are related to each other as

\begin{equation}
g_{a\gamma\gamma} \approx -\cot{\theta_w} g_{a\gamma Z} \approx -1.9 \, g_{a\gamma Z}.
\end{equation}

The corresponding (Barr-Zee) correction to $\left(g-2\right)_f$ is illustrated in Fig. \ref{fig:dbrz}, where the internal boson line represents either a photon or a $Z$ boson, and is given by \cite{Marciano:2016yhf,Bauer:2017ris,Liu:2022tqn},

\begin{align}\label{eq:BZ}
\begin{split}
& \Delta a^{BZ}_{\gamma} = \frac{m_f g_{aff}g_{a\gamma\gamma}}{8\pi^2} \ln{\Lambda^2} - \frac{m_fg_{aff}g_{a\gamma\gamma}}{8\pi^2} \\
& \times \int_0^1\int_0^x \left[ \left(1+3y\right)\ln{\Delta_{\gamma}}+\frac{m_f^2y^3}{\Delta_{\gamma}} \right] dydx,
\end{split} \\
& \Delta_{\gamma} = m_f^2y^2 + m_a^2\left(1-x\right),
\end{align}
in case of photon, while in case of the $Z$-boson one has to replace

\begin{align}
& \Delta_{\gamma} \to \Delta_Z =  m_f^2y^2 + m_a^2\left(1-x\right)+m_Z^2\left(x-y\right), \\
& g_{a\gamma\gamma } \to -\frac{4\sin^2{\theta_w}-1}{4\sin{\theta_w}\cos{\theta_w}}g_{a\gamma Z }.
\end{align}

The explicit dependence on the cut-off scale $\Lambda$ in these expressions indicates the incompleteness of the effective theory and the necessity of some specific UV completion to obtain the full, model-dependent result.

However, since our primary goal is to derive bounds on ALP interactions - rather than to compute the magnetic moment correction as such - the above expressions appear to be sufficient. This requires adopting a reasonable value for $\Lambda$, which, given the absence of new weak-scale states at LHC energies, is expected to be at least $\mathcal{O}\left(\text{TeV}\right)$. A benchmark value $\Lambda = 1 \, \text{TeV}$ is commonly used in the literature  \cite{Marciano:2016yhf,Bauer:2017nlg,Bauer:2017ris,Buen-Abad:2021fwq,Agrawal:2021dbo,Liu:2022tqn} for studying ALP constraints. For this choice, the logarithmically divergent term dominates the total ALP-photon loop contribution to $\left(g-2\right)_{\mu}$, accounting for roughly $90 \%$ of this effect in case of $m_a = 1 \, \text{GeV}$.

On the other hand, larger values of $\Lambda$ leads to further enhancement of the logarithmic term, providing a larger correction and, consequently, tighter constraints on the ALP-photon coupling \cite{Buen-Abad:2021fwq}. This makes $1 \, \text{TeV}$ a justified choice in order to obtain a conservative exclusion limit, and we adopt this value throughout our analysis.

It is also important to note that, unlike \ref{eq:yuk}, the expression \ref{eq:BZ} depends on the relative sign between $g_{aff}$ and $g_{a\gamma\gamma}$. The scenario $\left|g_{aff}\right| = \left|g_{a\gamma\gamma}\right|$ leads to weaker constraints, as the two contributing diagrams partially cancel each other.

Furthermore, it is evident that $Z$-boson constitutes only a minor correction, amounting to a few percent, relative to that of the photon.

\begin{figure*}
    \centering
    \includegraphics[width=0.31\linewidth]{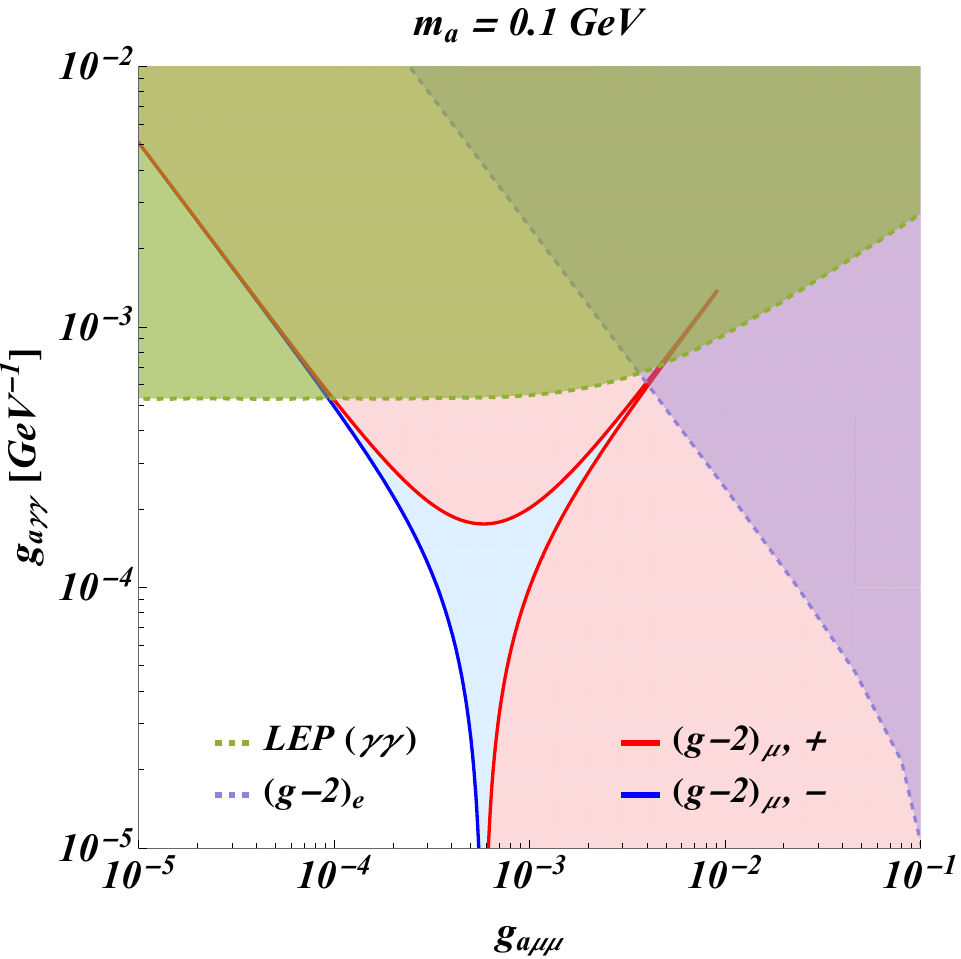}
    \includegraphics[width=0.31\linewidth]{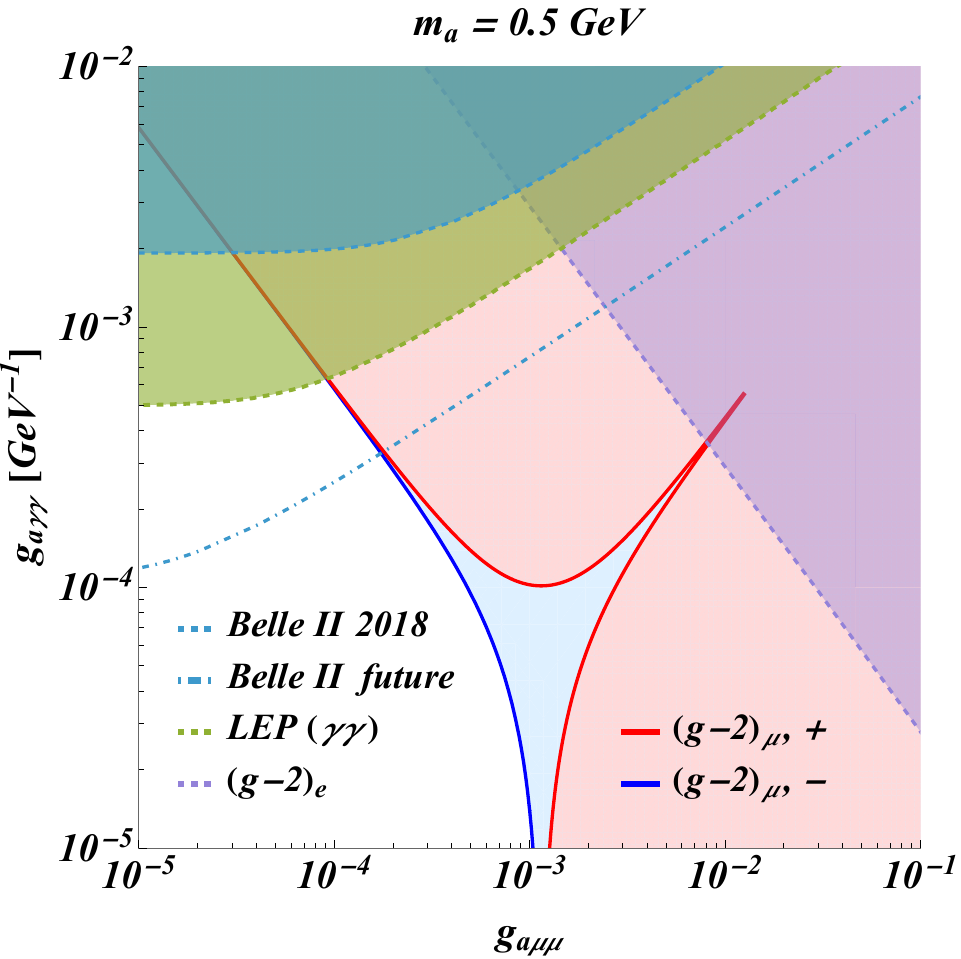}
    \includegraphics[width=0.31\linewidth]{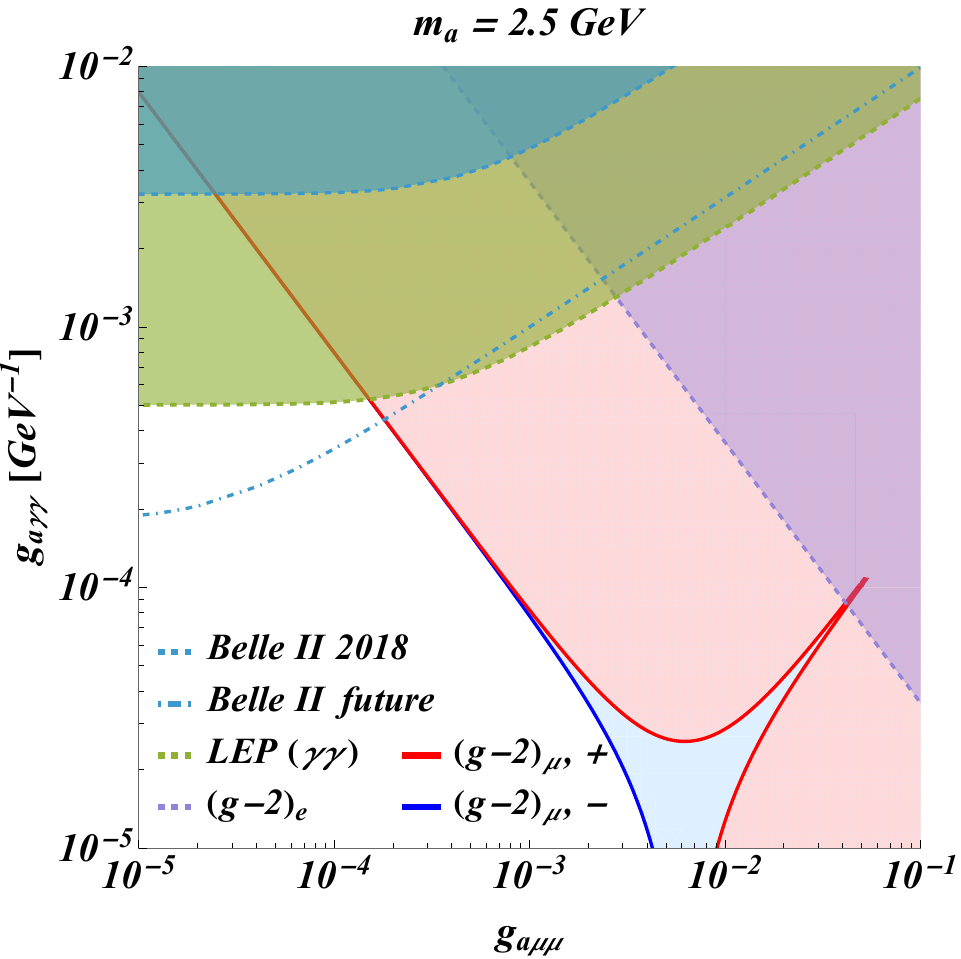}
    \caption{Coupling constraints, projected onto the  $\left(g_{a\mu\mu},g_{a\gamma\gamma}\right)$ plane (the colored regions are excluded at $2\sigma$ level). Red refers to $\left|g_{a\mu\mu}\right|=\left|g_{a\gamma\gamma}\right|$ scenario, blue stands for $\left|g_{a\mu\mu}\right|=-\left|g_{a\gamma\gamma}\right|$. The visible thin line in the first case represents the situation where the two contributions cancel each other out. Dashed lines are used for the model-dependent bounds. Dot-dashed line for Belle II is the projection based on the assumption of $50 \, \text{ab}^{-1}$ integrated luminosity.}
    \label{fig:LeptonBZ1ConstrALP}
\end{figure*}

\begin{figure*}
    \centering
    \includegraphics[width=0.45\linewidth]{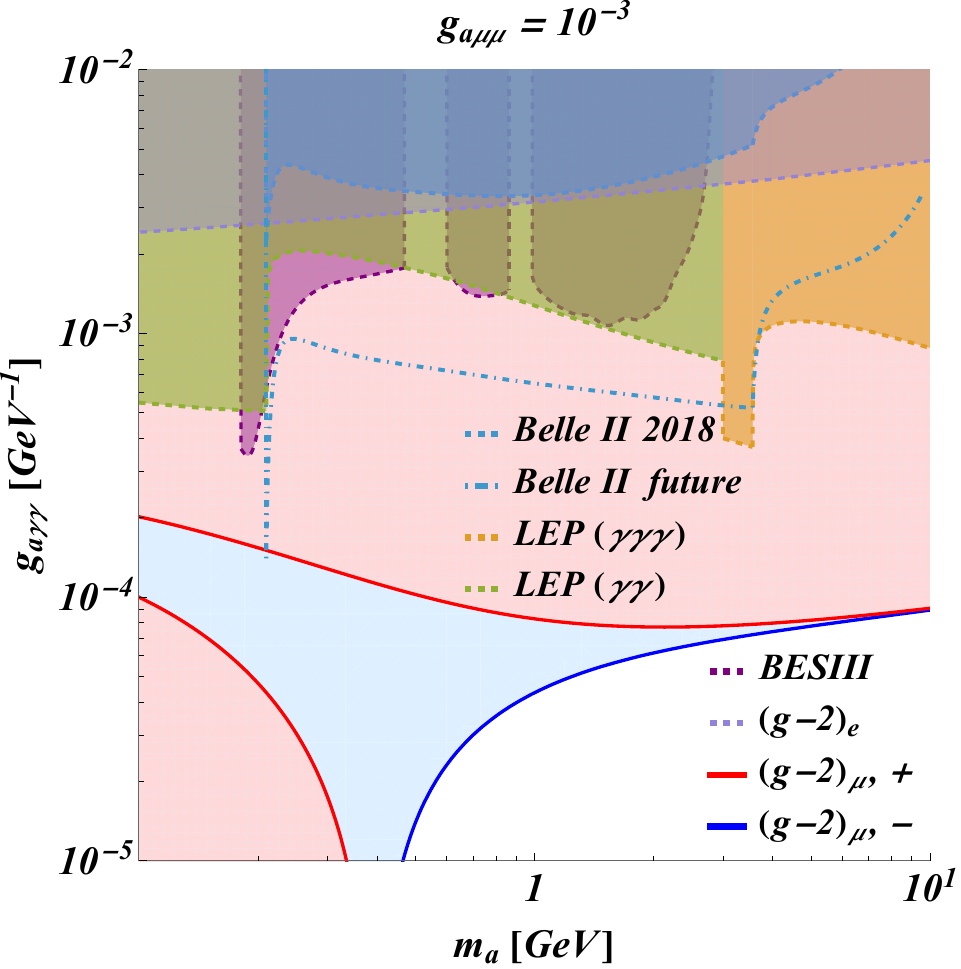}
    \includegraphics[width=0.45\linewidth]{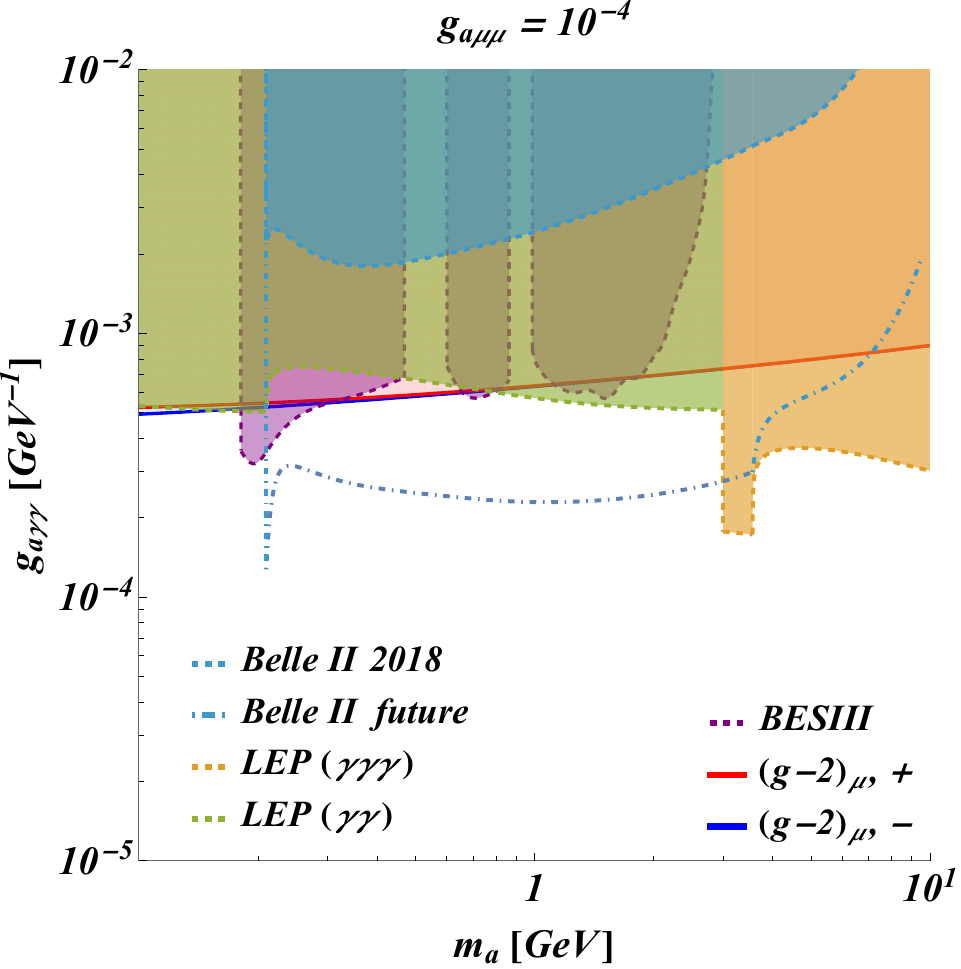}
    \caption{Coupling constraints, projected onto the  $\left(m_a,g_{a\gamma\gamma}\right)$ plane (the colored regions are excluded at $2\sigma$ level). Red refers to $\left|g_{a\mu\mu}\right|=\left|g_{a\gamma\gamma}\right|$ scenario, blue stands for $\left|g_{a\mu\mu}\right|=-\left|g_{a\gamma\gamma}\right|$. The visible thin line in the first case represents the situation where the two contributions cancel each other out. For electron the difference between two possibilities is negligible in this domain, as the Barr-Zee diagram dominates. It also becomes noncompetitive for lower values of $g_{a\mu\mu}$ and thus now shown on the right panel. Dashed lines are used for the model-dependent bounds. Dot-dashed line for Belle II is the projection based on the assumption of $50 \, \text{ab}^{-1}$ integrated luminosity.}
    \label{fig:LeptonBZ2ConstrALP}
\end{figure*}

Fig.~\ref{fig:LeptonBZ1ConstrALP}  presents our results for three different ALP masses, $m_a = 0.1$, $0.5$, and $2.5$ GeV, projected onto the $\left(g_{a\mu\mu}, g_{a\gamma\gamma}\right)$ plane. 
Furthermore in Fig.~\ref{fig:LeptonBZ2ConstrALP}, we also show two projections onto the $\left(m_a,g_{a\gamma\gamma}\right)$ plane for the values of ALP-muon coupling $g_{a\mu\mu} = 10^{-3}$ and $g_{a\mu\mu} = 10^{-4}$. 
We emphasize that, beyond the effective Lagrangians of Eqs.~(\ref{eq:der}) and (\ref{eq:ew}) - which represent the only dimension-5 operators allowed for ALPs based on general theoretical considerations - and the well-motivated choice of $\Lambda = 1 \, \text{TeV}$ explained above, these exclusion limits do not rely on any further model-specific assumptions\footnote[1]{It was also found that in case of the electron the constraints obtained in this way are much weaker and do not provide any improvement upon existing results.}. 

In addition to the $(g-2)_\mu$ constraints shown in Figs.~\ref{fig:LeptonBZ1ConstrALP} and \ref{fig:LeptonBZ2ConstrALP}, we also incorporate adapted bounds from $e^+e^-$ collider experiments and electron magnetic dipole moment \cite{Mimasu:2014nea,Bauer:2017ris,Dolan:2017osp,Liu:2022tqn,Pustyntsev:2024ygw,Belle-II:2020jti,BESIII:2024hdv,L3:1995nbq,L3:1994shn,Beacham:2019nyx,Fan:2022eto,Aoyama:2014sxa}. These constraints, however, are model-dependent, relying on the assumption of lepton-universal ALP couplings and visible decays into two-photon final states.

Nevertheless, we see from Figs.~\ref{fig:LeptonBZ1ConstrALP} and \ref{fig:LeptonBZ2ConstrALP} that the $e^+e^-$ collider experiments and magnetic moment measurements complement each other, allowing to explore the broader range of ALP parameters - where one method loses sensitivity, the other provides additional constraints, ensuring comprehensive coverage. When combined, they may serve as one of the most effective strategies for exploring the MeV-GeV mass range for BSM physics.

Finally, we also revised constraints which can derived for a generic scalar coupled to the SM via

\begin{equation}
\mathcal{L}= -g_{sll} s \, \Bar{l} l - \frac{g_{s \gamma \gamma}}{4}sF^{\mu \nu}F_{\mu \nu}- \frac{g_{s \gamma Z}}{2}sF^{\mu \nu}Z_{\mu \nu}.
\end{equation}

Couplings $g_{s \gamma \gamma}$ and $g_{s \gamma Z}$ obey the relation similar to \eqref{eq:BZ}. The resulting corrections to the magnetic moment are given by the expressions \cite{Pustyntsev:2024ygw}

\begin{align}
& \Delta a^Y = \frac{m_f^2g_{sff}^2 }{8 \pi^2}  \int_0^1  \frac{\left(1-x\right)^2\left(1+x\right)}{m^2_s x + m_f^2\left(1-x\right)^2} \, dx, \\
\begin{split}
& \Delta a^{BZ}_{\gamma} = \frac{m_f g_{sff}g_{s\gamma\gamma}}{8\pi^2} \ln{\Lambda^2} - \frac{m_fg_{sff}g_{s\gamma\gamma}}{4\pi^2} \\
& \times \int_0^1\int_0^x \ln{\Delta_{\gamma}} \, dxdy,
\end{split}
\end{align}
where the same notation is used as in the case of ALPs, except that $m_a$ is replaced with $m_s$ in $\Delta_{\gamma}$. One important modification is that the Yukawa-like term $\Delta a^Y$ is now positive\footnote[2]{This, in principle, opens the possibility of a scenario in which the scalar and pseudoscalar contributions cancel each other at the one-loop level - an effect explored, for example, in \cite{Balkin:2021rvh}.}.

\begin{figure}
\centering
\includegraphics[width=1\linewidth]{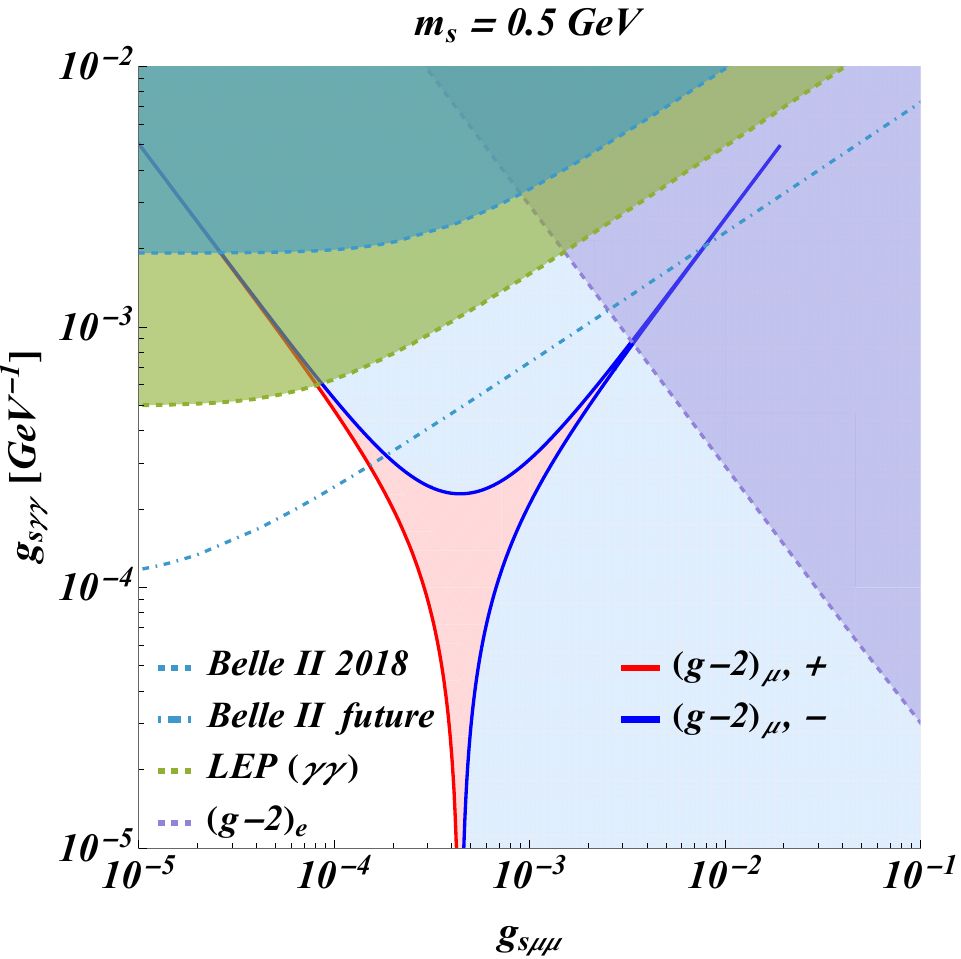} 
\caption{Coupling constraints, projected onto the  $\left(g_{s\mu\mu},g_{s\gamma\gamma}\right)$ plane (the colored regions are excluded at $2\sigma$ level). Red refers to $\left|g_{s\mu\mu}\right|=\left|g_{s\gamma\gamma}\right|$ scenario, blue stands for $\left|g_{s\mu\mu}\right|=-\left|g_{s\gamma\gamma}\right|$. The visible thin line in the second case represents the situation where the two contributions cancel each other out. Dashed lines are used for the model-dependent bounds. Dot-dashed line for Belle II is the projection based on the assumption of $50 \, \text{ab}^{-1}$ integrated luminosity.}
\label{fig:scalars}
\end{figure}

Our findings indicate that the constraints in this case are of comparable strength to those in the pseudoscalar particle scenario, as can be seen by comparison of Figs.~\ref{fig:LeptonConstr} and \ref{fig:scalars}. We assume that the fermion couplings $g_{sff}$ scale proportionally with the corresponding fermion masses $m_f$ \cite{Pustyntsev:2024ygw} - a choice motivated by models in which the scalar arises as a Nambu-Goldstone boson of some spontaneously broken high-scale symmetry with the derivative couplings naturally inherited\footnote[3]{Nevertheless, a variety of UV-complete models exist in which this statement may no longer hold \cite{Delaunay:2025lhl}; however, a detailed study of such scenarios lies beyond the scope of this work.}. Our main conclusions for the pseudoscalar case remain applicable to the scalar case as well.

\section{Dark photon contributions to $\left(g-2\right)_{\mu}$}\label{sec3}

The minimal dark photon model predicts a QED-like coupling between the massive vector field $A'$ and SM fermions of the form

\begin{equation}\label{eq:vector}
\mathcal{L} = -\epsilon e \Bar{f} \slashed{A}' f,
\end{equation}
with $\epsilon$ being the kinetic mixing parameter between the dark and SM photons. The 1-loop correction to the magnetic moment is analogous to previously considered Yukawa-like diagram \ref{fig:dyuk} and is expressed via \cite{Carlson:2012pc}

\begin{equation}
\Delta a = \frac{\alpha \epsilon^2m_f^2}{\pi} \int_0^1 \frac{x^2\left(1-x\right)}{m_f^2x^2 + m_A^2\left(1-x\right)}dx,
\end{equation}
where $m_A$ denotes the dark photon mass.

It is straightforward to also consider the pseudovector mediator

\begin{equation}\label{eq:axial}
\mathcal{L} = -\epsilon e \Bar{f} \gamma^5\slashed{A}'  f,
\end{equation}
which contributes to the magnetic moment as

\begin{equation}
\Delta a = -\frac{\alpha \epsilon^2m_f^2}{\pi} \int_0^1 \frac{x\left(1-x\right)\left(4-x\right)+2\frac{m_f^2}{m_A^2}x^3}{m_f^2x^2 + m_A^2\left(1-x\right)}dx.
\end{equation}

\begin{figure}
\centering
\includegraphics[width=1\linewidth]{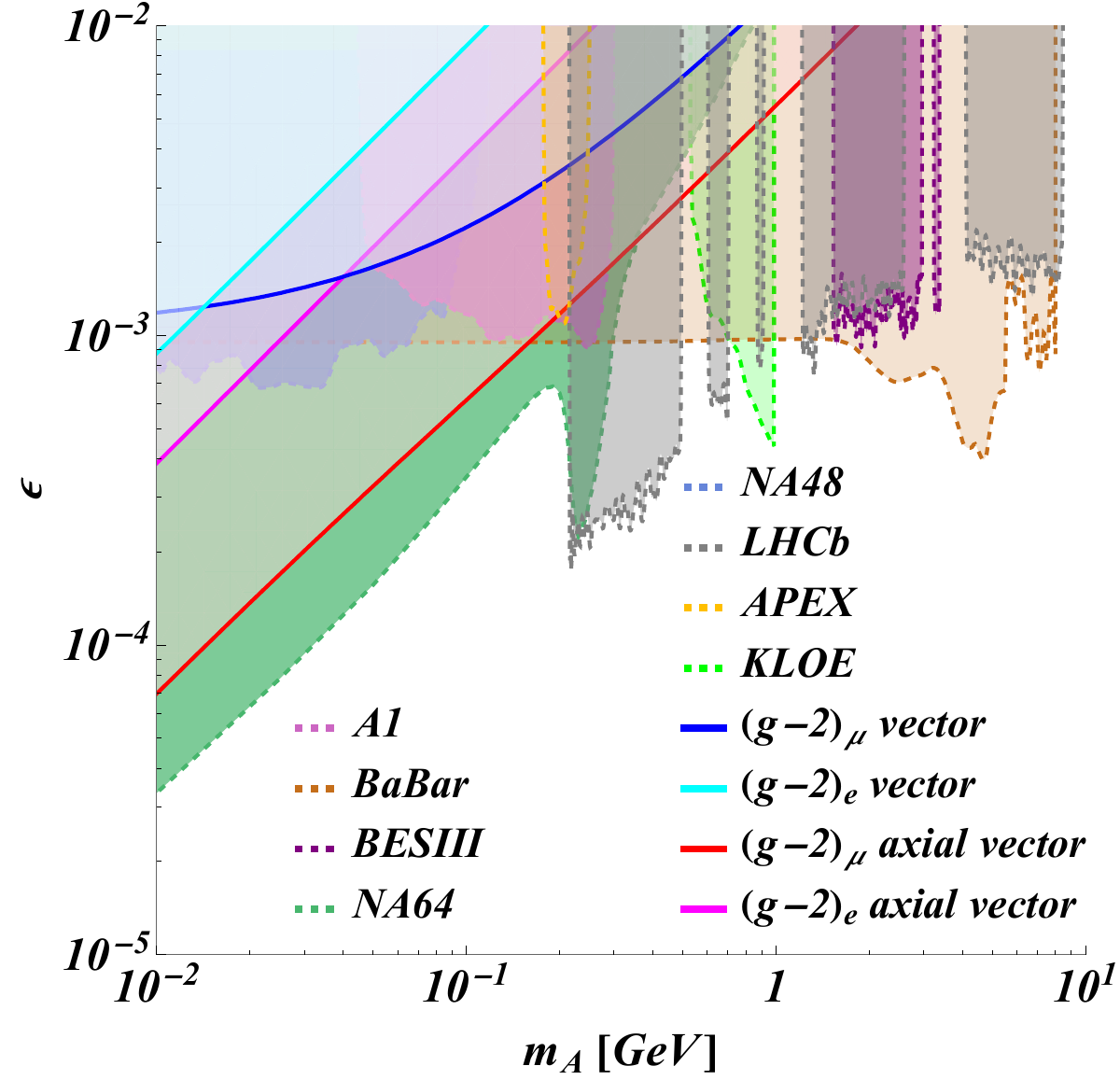} 
\caption{Regions of the $A'$ parameter space excluded by $\left(g-2\right)_{l}$, collider and fixed-target experiments (dashed lines) at $2\sigma$ level.}
\label{fig:vectors}
\end{figure}

The crucial difference between the two is that, in the vector case, the  $k^{\alpha}k^{\beta}/m_A^2$ term in the particle propagator vanish due to the current conservation, while in the pseudovector case it's contribution is proportional to $m_f^2/m_A^2$ and clearly, becomes more important for lower-mass axial vectors. This is reflected in Fig.~\ref{fig:vectors}, which demonstrates that axial vector constraints are more stringent in the low mass region compared to the true vector ones. 

Bounds from collider and fixed-target experiments \cite{APEX:2011dww,Merkel:2014avp,NA482:2015wmo,KLOE-2:2016ydq,BESIII:2017fwv,LHCb:2019vmc,NA64:2023wbi} are also shown. It is evident from Fig.~\ref{fig:vectors} that the parameter space sensitive to $\left(g-2\right)_{l}$ is already entirely covered by the NA64 exclusion limit. Nevertheless, we argue that the obtained result remains relevant, as the NA64 bound is model-dependent, relying on the assumption of an invisible decay into dark fermions with masses of $m_A/3$, and a specific coupling to the dark photon, $\alpha_D = 0.1$. In contrast, the $\left(g-2\right)_{l}$ exclusion limit is derived solely from generic dimension-4 couplings of Eqs.~(\ref{eq:vector}) and (\ref{eq:axial}), without any additional model assumptions.

\section{Conclusion}\label{sec4}

Over the past few years, the muon magnetic anomaly has been one of the most widely discussed topics in the field of particle physics. Significant advancements in both experimental measurements and theoretical calculations were made, which ultimately allowed the tensions to be resolved within the Standard Model.

At the same time, one of the most powerful tools for constraining possible BSM scenarios was obtained, which we used to revise the parameter space of potential scalars, pseudoscalars, vectors and axial vectors in the MeV-GeV mass range. This domain was chosen as the existing constraints therein remain relatively underexplored.

No specific UV-complete model was assumed in deriving bounds on scalar and pseudoscalar interactions. The UV cut-off $\Lambda = 1 \, \text{TeV}$ was selected to provide a conservative estimate of these limits. Depending on the value of $g_{a\mu\mu}$, the resulting limits can be up to an order of magnitude more restrictive than those previously studied, clearly illustrating the usefulness of this analysis.

It can also be seen that lepton magnetic moment measurements and collider searches complement one another by probing regions of the parameter space where different sensitivities are observed. With ongoing data collection at BESIII and Belle II, there is significant potential to further tighten the constraints on these scenarios.

Finally, for vector and axial-vector cases, although the constraints obtained are less stringent and do not establish entirely new bounds, they nevertheless offer competitive exclusion limits that remain independent of other search strategies and, in contrast to NA64 limits, do not imply specific assumptions about the $A'$ decays.

\section*{Acknowledgments}
This work was supported by the Deutsche Forschungsgemeinschaft (DFG, German Research Foundation), in part through the Research Unit [Photon-photon interactions in the Standard Model and beyond, Projektnummer 458854507 - FOR 5327], and in part through the Cluster of Excellence [Precision Physics, Fundamental Interactions, and Structure of Matter] (PRISMA$^+$ EXC 2118/1) within the German Excellence Strategy (Project ID 39083149).

\bibliographystyle{apsrev4-1} 
\bibliography{bibliography}% common bib file

\end{document}